\documentclass[12pt]{article}
\usepackage{epsfig}
\def\makeatletter{\catcode`\@=11}% 11:letter
\makeatletter
\def\mathbox#1{\hbox{$\m@th#1$}}%
\def\math@ccstyles#1#2#3#4#5#6#7{{\leavevmode
      \setbox0\mathbox{#6#7}%
      \setbox2\mathbox{#4#5}%
      \dimen@ #3%
      \baselineskip\z@\lineskiplimit#1\lineskip\z@
      \vbox{\ialign{##\crcr
             \hfil \kern #2\box2 \hfil\crcr
             \noalign{\kern\dimen@}%
             \hfil\box0\hfil\crcr}}}}
\def\mathaccstyles{\math@ccstyles\maxdimen}
\def\maththroughstyles{\math@ccstyles{-\maxdimen}}
\def\unitmatrixDT%
 {\maththroughstyles{.45\ht0}\z@\displaystyle {\mathchar"006C}\displaystyle 1}

\def\tfrac#1#2{{\textstyle{#1\over #2}}}

\def\two{\tfrac{1}{2}}

\begin{document}
\renewcommand{\arraystretch}{1.5}

\rightline{UG-10/98}
\rightline{hep-th/9806069}
\vspace{3truecm}
\centerline{\bf \large On M-9-branes}
\vspace{1cm}
%authors
\centerline{Eric~Bergshoeff and Jan~Pieter~van der Schaar}
\vspace{0.6truecm}
\centerline{\it Institute for Theoretical Physics}
\centerline{\it Nijenborgh 4, 9747 AG Groningen}
\centerline{\it The Netherlands}
\vspace{2truecm}
\centerline{ABSTRACT}
\vspace{.5truecm}

We discuss some properties of the conjectured M-9-brane. We investigate
both the worldvolume action as well as the target space solution.
The worldvolume action is given by a gauged sigma model which, via
dimensional reduction and duality, is shown to be related to the
worldvolume actions of the branes of ten-dimensional superstring theory.
The effective tension of the M-9-brane
scales as $(R_{11})^3$,
where $R_{11}$ is the radius of an $S^1$--isometry direction.
This isometry enables us to add a cosmological constant
to eleven-dimensional supergravity.

The target space solution corresponding to the M-9-brane is a (wrapped)
domain wall
solution of massive eleven--dimensional supergravity. 
This solution breaks half of the bulk supersymmetry. 
We consider both single M-9-branes as well as a system of two M-9-branes.
In both cases one can define regions in spacetime, separated by the
domain walls, with zero cosmological constant. In these regions
the limit $R_{11} \rightarrow \infty$ can be taken in which case
the M-9-brane is unwrapped and a massless
eleven-dimensional supergravity theory is obtained.
\newpage

\noindent{\bf 1.\ Introduction}
\vspace{0.5cm}

Given the fact that eleven-dimensional supergravity arises in the 
strong-coupling limit of Type IIA superstring theory \cite{witten}, 
one would expect
that all Type IIA branes arise as solutions of the equations
of motion corresponding to eleven--dimensional supergravity \cite{paul}. 
This is indeed the
case for all known branes of Type IIA superstring theory except for the
D-8-brane which arises as a domain wall solution in ten dimensions 
\cite{PW,berogrpato1} and whose eleven-dimensional origin so far has
remained unclear (see Figure 1). It has been conjectured that the
eleven-dimensional origin of the D-8-brane is an M-9-brane\footnote{
For earlier discussions
on the conjectured M-9-brane, see \cite{berogrpato1,gy}.}.
In this work we investigate some properties of this conjectured M-9-brane.

One indication that indeed an M-9-brane exists in eleven dimensions
comes from a study of the M-theory superalgebra \cite{townsend,hull}:
\vspace{1mm}

\begin{equation}
\{Q_\alpha, Q_\beta\} = (\Gamma^MC)_{\alpha\beta}P_M + {1\over 2}
(\Gamma_{MN}C)_{\alpha\beta}Z^{MN} + {1\over 5!}(\Gamma_{MNPQR})_{\alpha\beta}
Y^{MNPQR}\, .
\end{equation}
\vspace{1mm}

\noindent The time component of the 2--form central charge $Z$ suggests 
a 9--brane which breaks half of the bulk supersymmetry according to
($\eta$ is the bulk supersymmetry parameter)

\begin{equation}
(1-\Gamma_{0123456789}) \eta = 0\, .
\label{susycondition}
\end{equation}
>From a 10-dimensional point of view (reducing over $x^{10}$)
this is exactly the chirality condition 

\begin{equation}
(1-\Gamma_{\star})\eta = 0\, ,
\end{equation}
with $\Gamma_\star = \Gamma_{10} = \Gamma_0\Gamma_1\cdots \Gamma_9$.
Hence the worldvolume theory on the 9-brane has 
$N=1$ chiral supersymmetry. This 10-dimensional worldvolume theory occurs
in the Ho{\u r}ava-Witten description of the $E_8 \times E_8$
heterotic string \cite{howi}.

It is nontrivial to find a solution of eleven-dimensional supergravity that
describes the expected M-9-brane. The double dimensional reduction
of such an M-9-brane must lead to the D-8-brane. However, the D-8-brane
solution arises in massive IIA supergravity \cite{romans} 
which is not directly obtainable by reduction of d=11 
supergravity. The reason for this is that massive IIA supergravity
contains a cosmological constant proportional to $m^2$ where $m$ is a
parameter with the dimension of a mass\footnote{The parameter $m$ can be
positive or negative. The sign of the cosmological constant is determined
by T-duality (see below) and is such that
for constant dilaton $\phi = \phi_0$ the vacuum solution  
to the Einstein equations is anti-de Sitter spacetime.}:

\begin{equation}
S_{\rm massive\ IIA}(m) \sim \int d^{10}x {\sqrt {|g|}}\left \{
e^{-2\phi} \left[ R-4(\partial \phi)^2 \right] + \two m^2 + \cdots \right\}\, .
\label{massiveIIA}
\end{equation}
There does not exist a corresponding massive extension of
the M--theory d=11 supergravity theory (for a recent discussion of this point,
see \cite{nogo}). There is a simple explanation for this.
The only massive extension one can write down reads

\begin{equation}
S_{\rm massive\ d=11} \sim \int d^{11}x {\sqrt {|g|}}\left \{
R + \two m^2 + \cdots \right\}\, .
\label{massived=11}
\end{equation}
However, the reduction of the cosmological constant
in the above action
leads to a cosmological constant in d=10 whose dilaton
coupling (in string--frame) does not vanish, as in (\ref{massiveIIA}).

\begin{figure} \label{fig1}
\begin{center}
\includegraphics[angle=0, width=150mm]{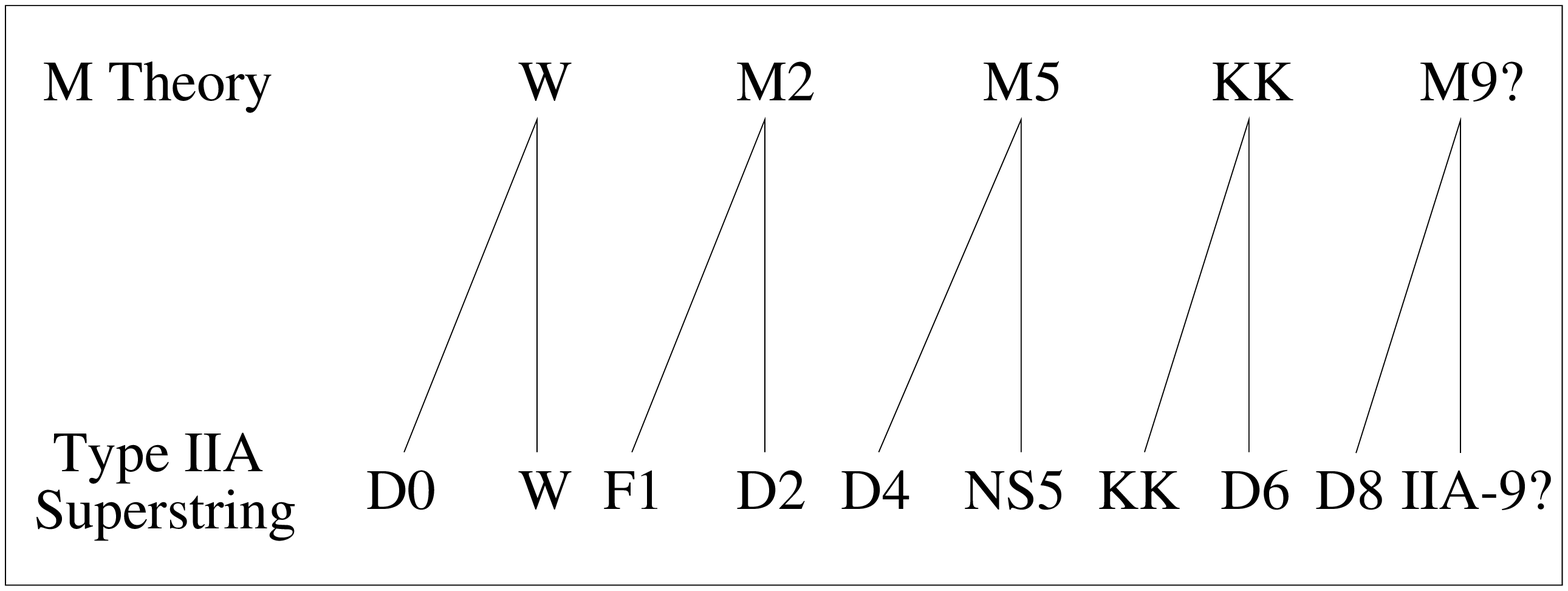}
\end{center}
Figure 1: {\small {\bf 
Relation between Type IIA branes and M-theory solutions:}
 Vertical lines imply direct dimensional reduction, diagonal
 lines double dimensional reduction. Besides Type IIA D-p--branes (p=0,2,4,6,8)
and M-p--branes (p=2,5), the figure contains waves (W), the fundamental
string (F1), Kaluza-Klein monopoles (KK) and a (conjectured) IIA-9-brane. 
The oxidation of the D-8-brane leads to the M-9-brane discussed in the text.}
\end{figure}

One way to avoid this obstruction is to consider a d=11 supergravity
background with an isometry generated by a (spacelike) Killing vector 
$k^\mu$ \cite{beloor}. This possibility was originally motivated by 
considering the M-theory origin of the D-2-brane in a massive background
\cite{bejaor,y,t}. As is well-known, the worldvolume action of the
massless D-2-brane can be rewritten as a massless M-2-brane action 
after a worldvolume Poincar\'e dualization of the Born-Infeld (BI) 1-form $V$
into an embedding coordinate $X$ \cite{paul2}:

\begin{equation}
\label{dr1}
dX = {}^\star dV\, .
\end{equation} 
In a massive background the D-2-brane action obtains an extra Chern-Simons
term proportional to $m$ \cite{BR,ght} and, on-shell, the duality relation
(\ref{dr1}) gets replaced by

\begin{equation}
\label{dr2}
\bigl (dX - mV\bigr) = {}^\star dV\, .
\end{equation}
It has been shown \cite{y,t} that in order to perform the duality
transformation off-shell one must introduce an auxiliary worldvolume
1-form A and replace the Chern-Simons term by\footnote{Note that on-shell 
$dA=dV$ and the two formulations of the Chern-Simons term coincide.}

\begin{equation}
mVdV \rightarrow m\bigl ( {\textstyle{{1\over 2}}} AdA - AdV\bigr )\, .
\end{equation}
The dualization of the BI 1-form $V$ into an embedding coordinate $X$
then leads to a massive M-2-brane action which is gauged sigma model. The
action is obtained from the massless M-2-brane action by replacing
the worldvolume derivatives by covariant derivatives:

\begin{equation}
d X^\mu \rightarrow  d X^\mu - Ak^\mu\, ,
\end{equation}
where $A$ is the auxiliary 1-form introduced above and 
$k^\mu$ is a Killing vector. Thus the construction of the massive 
M-2-brane action requires the existence of a Killing vector. 

In view of the above discussion it is natural to
assume the existence of the same Killing vector $k^\mu$ in order to
construct an eleven-dimensional supergravity theory with a
cosmological constant. Given this Killing vector  one can define an extra 
scalar\footnote{We use a mostly minus signature $(+,-, \cdots ,-)$ so that
$|k|^2$, as defined below, is 
positive for a spacelike Killing vector $k^\mu$. We work in units in which 
the eleven-dimensional Planck-length is equal to one.} 

\begin{equation}
|k|^2 = - k^\mu k^\nu g_{\mu\nu} = (R_{11})^2\, .
\end{equation}
Using this scalar one can modify the Lagrangian (\ref{massived=11}) such 
that upon 
reduction it leads to a cosmological constant with the correct dilaton
coupling:

\begin{equation}
S_{\rm massive\ M} \sim \int d^{11}x {\sqrt {|g|}}\left \{
R + \two |k|^4 m^2 + \cdots \right\}\, .
\label{massiveMm}
\end{equation}
This leads to the definition of ``massive d=11 supergravity''. The bosonic
part of the action has been given in \cite{beloor}.

The massive supergravity theory defined above is not a proper
d=11 supergravity theory in the sense that it is only defined
for backgrounds with a vanishing Lie derivative. {\it A priory} we cannot
exclude the possibility that such backgrounds are special solutions of a
yet to be constructed proper d=11 supergravity theory with a cosmological
constant. Given the no-go theorem of \cite{nogo} this seems unlikely to be
the case. We therefore continue our analysis using the massive
d=11 supergravity theory defined above.
In fact, as we will show in this work,
one can relax the restriction of vanishing Lie derivative to be valid only in
certain regions of the d=11
spacetime which are separated by domain wall M-9-brane solutions.
For this to work we must use the fact that the cosmological constant
is not necessarily constant everywhere
but can be taken to be {\it piecewise constant} \cite{PW,berogrpato1}. 
In the case of massive IIA supergravity
this is seen by using a
formulation where the mass parameter $m$ is replaced by a 9-form
potential $C^{(9)}$
with curvature $G^{(10)}$ \cite{berogrpato1}\footnote{In the case of massive 
d=11 supergravity one must replace $m$ by a 10-form potential.}:

\begin{equation}
S_{\rm massive\ IIA}(C^{(9)}) \sim \int d^{10}x {\sqrt {|g|}}\left \{
e^{-2\phi} \left[ R-4(\partial \phi)^2 \right] - \frac{1}{2 \times 10!}
(G^{(10)})^2 
+ \cdots \right\}\, .
\label{C9}
\end{equation}
This 9-form potential naturally appears in the
low-energy limit of Type IIA superstring theory \cite{pol}.
The sign of the kinetic term for $C^{(9)}$ follows, via T-duality, from
the (standard) sign of the kinetic term for the other Ramond-Ramond
p-form potentials $C^{(p)}\ (p<9)$. This in turn fixes the
sign of the cosmological constant in (\ref{massiveIIA}).
More precisely, the equation of motion of $C^{(9)}$

\begin{equation}
\label{eomc9}
d\ {}^* G^{(10)} = 0
\end{equation}
is solved for by

\begin{equation}
\label{solc9}
G^{\mu_1 \ldots \mu_{10}}(C)=\frac{1}{\sqrt{|g|}}
\epsilon^{\mu_1 \ldots \mu_{10}} c \, ,
\label{pdual}
\end{equation}
where $c$ is an integration constant. Comparing with (\ref{massiveIIA})
we find that $c=\pm m$.

The expression of $G^{(10)}$ given in (\ref{solc9}) is not the most general 
solution of (\ref{eomc9}). In the presence of a domain wall, 
the integration constant $c$ can be {\it piecewise
constant} \cite{berogrpato1}. This possibility is excluded in the
formulation without the 9-form potential where the parameter $m$ is
constant everywhere. 
In particular, the presence of a domain wall allows us to
define the region of d=11 spacetime at one side of the domain wall to
have zero cosmological constant, i.e.~$m=0$. This observation will be of
use later.

In this work we will use the massive d=11 supergravity theory to investigate
the oxidation of the D-8-brane target space solution into an
M-9-brane target space solution.
The organization of this paper is as follows.
In section 2 we first discuss some properties of the
M-9-brane worldvolume action. In particular, we discuss how the
tension scales with $R_{11}$ and how the leading Namnu-Goto term of the
M-9-brane worldvolume action is related,
via reduction and duality, to the worldvolume actions of the branes in 
ten-dimensional string theory.
In section 3 we discuss some properties of the D-8-brane target space 
solution. 
In section 4 we investigate the corresponding M-9-brane target space
 solution. We discuss both the
single M-9-brane as well as the two domain wall system. The supersymmetry
properties of the D-8-brane and M-9-brane are discussed in section 5.
Finally, in section 6 we present our conclusions.

\vspace{0.5cm}
\noindent{\bf 2.\ The M-9-brane worldvolume action}
\vspace{0.5cm}

One way to see that it is nontrivial to construct the worldvolume action
for an  M-9-brane moving in eleven {\it decompactified} dimensions is
to consider the scaling of the tension with $R_{11}$. It is instructive
to do this analysis for all the branes of M-theory (see Figure 1).
Our starting point is the gauged $\sigma$-model approach. This approach
enables us to reformulate 
$p$-branes moving in a d=10 target spacetime as 
$p$-branes moving in a d=10 submanifold of a d=11 target spacetime.
In the latter case the tension of the $p$-brane is measured in terms of d=11
quantities.
The gauged $\sigma$-model plays a crucial role in the construction of the
worldvolume action of the Kaluza-Klein (KK) monopole \cite{bejaor}.

Consider the Nambu-Goto part of a general Type IIA p-brane in ten
dimensions with dilaton coupling parameter $\alpha$:

\begin{equation}
S (d=10) = -T \int d^{p+1}x \hspace{4pt} e^{\alpha\phi} 
\sqrt{\mid \partial_i X^{\mu} \partial_j X^{\nu} g_{\mu\nu} \mid}\, .
\label{action2}
\end{equation}
This action describes the dynamics of a p-brane moving in ten dimensions. 
We next write down an equivalent action describing the same p-brane but 
now moving in a ten-dimensional subspace  of an eleven-dimensional manifold
with an isometry generated by a spacelike Killing vector 
${\hat k}^{\hat\mu}$ (we indicate eleven-dimensional
fields and indices with a hat):

\begin{equation}
S (d=11) =-T \int d^{p+1}\hat{x} \hspace{4pt} |\hat{k}|^{\beta} 
\sqrt{\mid \partial_i \hat{X}^{\hat{\mu}} \partial_j \hat{X}^{\hat{\nu}}
\hat{\Pi}_{\hat{\mu}\hat{\nu}} \mid}\, .
\label{action1}
\end{equation}
The projector

\begin{equation}
\hat{\Pi}_{\hat{\mu}\hat{\nu}}=\hat{g}_{\hat{\mu}\hat{\nu}}+|\hat{k}|^{-2}
\hat{k}_{\hat{\mu}}\hat{k}_{\hat{\nu}}
\end{equation}
restricts the dynamics of the brane to the space 
orthogonal to the isometry direction. 
The effective tension scales as $|{\hat k}|^\beta = (R_{11})^{\beta}$ 
where $\beta$ is
a scaling parameter to be determined below.
Note that the d=11 background metric has vanishing Lie derivative with
respect to the Killing vector ${\hat k}$.

The two actions (\ref{action2}) and (\ref{action1}) are related to each
other via a direct dimensional reduction over the isometry direction.
To perform this reduction it is convenient to use coordinates $\hat \mu =
(\mu, z)$ adapted
to the isometry. In this basis the only
non-vanishing components of the projector $\hat{\Pi}_{\hat{\mu}\hat{\nu}}$  are
\begin{equation}
\hat{\Pi}_{\mu\nu}=|\hat{k}|^{-1} g_{\mu\nu}\, .
\end{equation} 
Using the relation 

\begin{equation}
\hat{g}_{zz}=-|\hat{k}|^2=-e^{\tfrac{4}{3}\phi}
\end{equation}
we find that the dilaton coupling parameter $\alpha$ of the ten-dimensional 
p-brane
and the scaling parameter $\beta$ of the corresponding eleven-dimensional 
p-brane are related to each other as follows:
\vskip .1truecm

\begin{equation}
\alpha = -\tfrac{1}{3}(p+1)+\tfrac{2}{3}\beta\, .
\label{eqab}
\end{equation}
\vskip .3truecm

For the standard branes of Type IIA superstring theory the dilaton
coupling parameter $\alpha$ takes on three different values (the discussion
below is summarized in Figure 2)\footnote{We do not consider here
purely gravitational branes like waves and monopoles. The KK 
monopole will be discussed below.}:

\begin{itemize}
\item
$\alpha=0$ for fundamental objects like the F1 string.

\item
$\alpha=-1$ for the D-p-branes (p=0,2,4,6,8).

\item
$\alpha=-2$ for solitonic objects like the NS5 brane.
\end{itemize}

\begin{figure} \label{fig2}
\begin{center}
\includegraphics[angle=0, width=150mm]{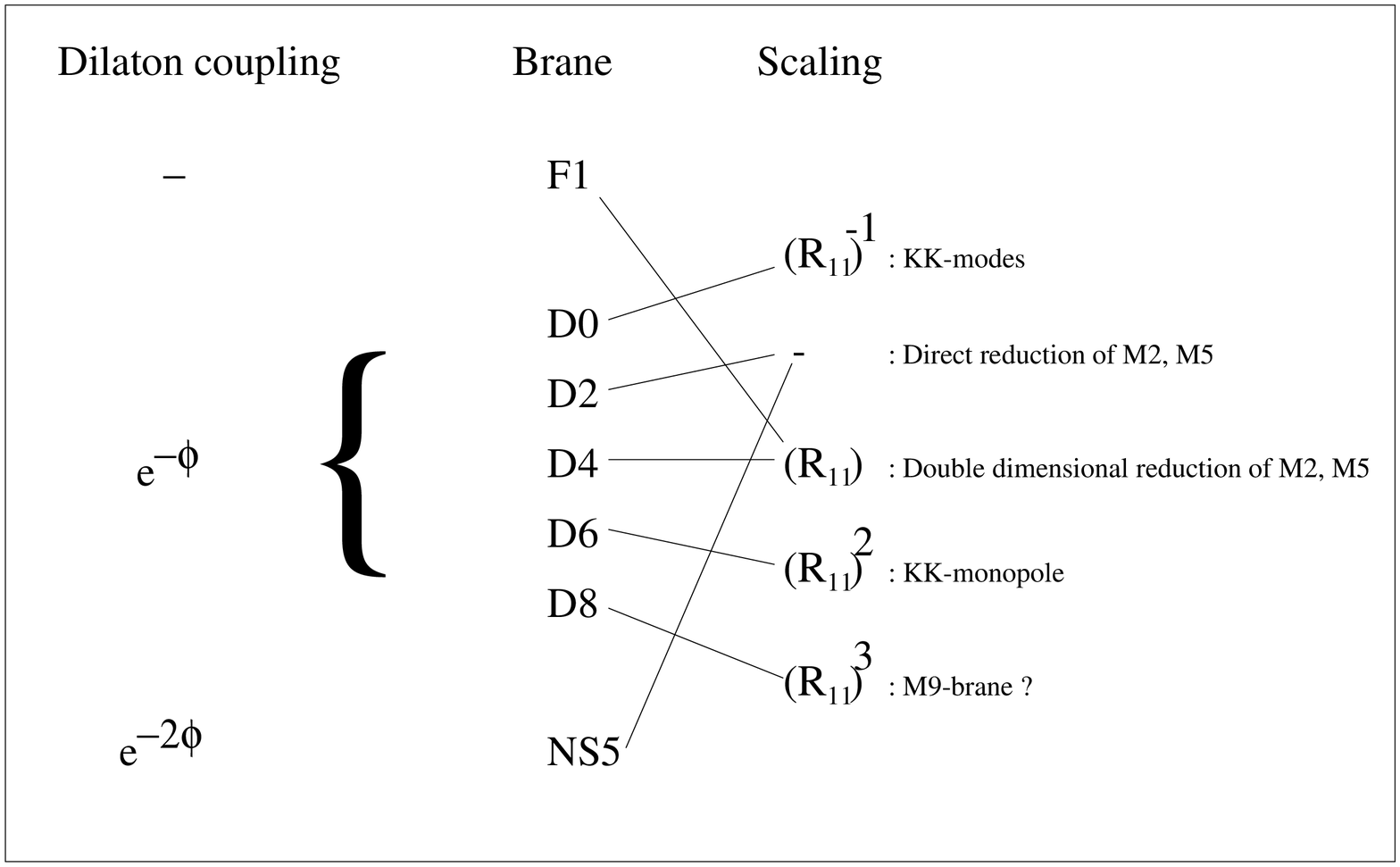}
\end{center}
Figure 2: {\small {\bf The scaling of Type IIA branes:} The Type IIA-branes
are divided into three classes according to their dilaton coupling. 
The lines in the figure indicate for each brane how the corresponding
M-brane tension scales with $R_{11}$.}
\end{figure}

\noindent This leads to the following five different values of 
the scaling parameter $\beta$:

\begin{itemize}
\item
$\beta=-1$: this is the appropriate scaling for the KK modes 
coming from $d=11$. In $d=10$ they give rise to the D-0-branes.

\item
$\beta=0$: In this case we are dealing with the D-2-brane and NS5-brane
which are related via a direct dimensional reduction to the M-2-brane
and M-5-brane, respectively. In these cases the tension does not scale with
$R_{11}$ and no extra worldvolume direction is developed\footnote{In order to 
rewrite the D-2-brane (NS5-brane) action as an M-2-brane (M-5-brane) action
one must dualize the BI 1-form into an embedding coordinate (D-2-brane)
or use the extra worldvolume scalar (NS5-brane).}.

\item
$\beta=1$: This is what we expect from the F1-string and D-4-brane
since they follow via a double dimensional reduction from the M-2-brane
and M-5-brane, respectively. The scaling of the tension with $R_{11}$ 
signatures the development of an extra worldvolume direction. Absorbing
the $R_{11}$-factor into the Nambu-Goto part, the gauged sigma model
action can be rewritten as the usual worldvolume action of the M-2-brane 
or M-5-brane\footnote{Note that in the case of the D-4-brane/M-5-brane action
the BI 1-form must be oxidized to a self-dual 2-form in order not to upset
the counting of the bosonic degrees of freedom.}.

\item
$\beta=2$: This happens for the D-6-brane. The D-6-brane can be 
obtained from the KK monopole in $d=11$ when reduced over the
U(1) isometry direction in the transversal Taub-NUT space.
This isometry direction can not be interpreted
as a worldvolume direction \cite{hull}. Since the monopole cannot move 
in the isometry direction the corresponding scalar has to be gauged away 
explicitly. This leads to a gauged $\sigma$-model worldvolume action
for the KK monopole \cite{bejaor}. 
The scaling of the M-theory monopole tension with $(R_{11})^2$ shows
that the isometry direction must be compact: the D-6-brane
cannot be oxidized to a brane moving in eleven decompactified dimensions.

\item
$\beta=3$ \cite{ropasc,giveon}: This case corresponds to the D-8-brane studied 
in this work.
We see that the M-9-brane, if it exists, cannot live in a decompactified 
d=11 manifold, but, like the M-theory monopole it needs a compact isometry
direction. 
\end{itemize}  

Before discussing the M-9-brane worldvolume action
it is instructive to first consider the
KK monopole. Since the d=10 KK monopole is T-dual to the Type IIB NS5-brane
it is a solitonic object whose effective tension scales with $\alpha = -2$.
Applying (\ref{eqab}) for $\alpha = -2$ and $p=5$ gives a scaling
parameter $\beta = 0$. Since the d=10 KK monopole 
is related to the d=11 KK monopole via a double dimensional reduction  
we expect to obtain $\beta=1$ and not $\beta = 0$. 
The resolution to this apparent puzzle
lies in the fact that the d=10 KK monopole action is not of the
standard form (\ref{action2}). Instead it belongs to the more general class
of worldvolume actions \cite{bejaor}
\begin{equation}
S (d=10) = -T \int d^{p+1}x \hspace{4pt} e^{\alpha\phi} |k^\prime|^\gamma
\sqrt{\mid \partial_i X^{\mu} \partial_j X^{\nu} g_{\mu\nu} \mid}
\label{kaction}
\end{equation}
with  $p=5, \alpha = -2$ and $\gamma = 2$. Here $k^{\prime \mu} 
\ne k^\mu $ is a Killing vector. 
In the case of the KK monopole
the Killing vector $k^{\prime \mu}$ refers to the Taub-NUT isometry direction.
As before, 
the action (\ref{kaction}) can be rewritten, for any value of $\gamma$,
as a gauged sigma model with an eleven-dimensional target space:

\begin{equation}
S (d=11) =-T \int d^{p+1}\hat{x} \hspace{4pt} |\hat{k}|^{\beta} 
|\hat{k}^\prime|^{\gamma} 
\sqrt{\mid \partial_i \hat{X}^{\hat{\mu}} \partial_j \hat{X}^{\hat{\nu}}
\hat{\Pi}_{\hat{\mu}\hat{\nu}} \mid}\, .
\end{equation}
Equation (\ref{eqab}) (corresponding to the special case $\gamma=0$) 
gets replaced by the more general formula:
\begin{equation}
\alpha = -\tfrac{1}{3}(p+1)+\tfrac{2}{3}(\beta-\two \gamma)\, .
\label{eqab2}
\end{equation}
Substituting the appropriate values for for $p,\alpha$ and $\gamma$
corresponding to the d=10 KK-monopole we obtain 
$\beta=1$, as expected.

\begin{figure} \label{actionkk}
\begin{center}
\includegraphics[angle=0, width=150mm]{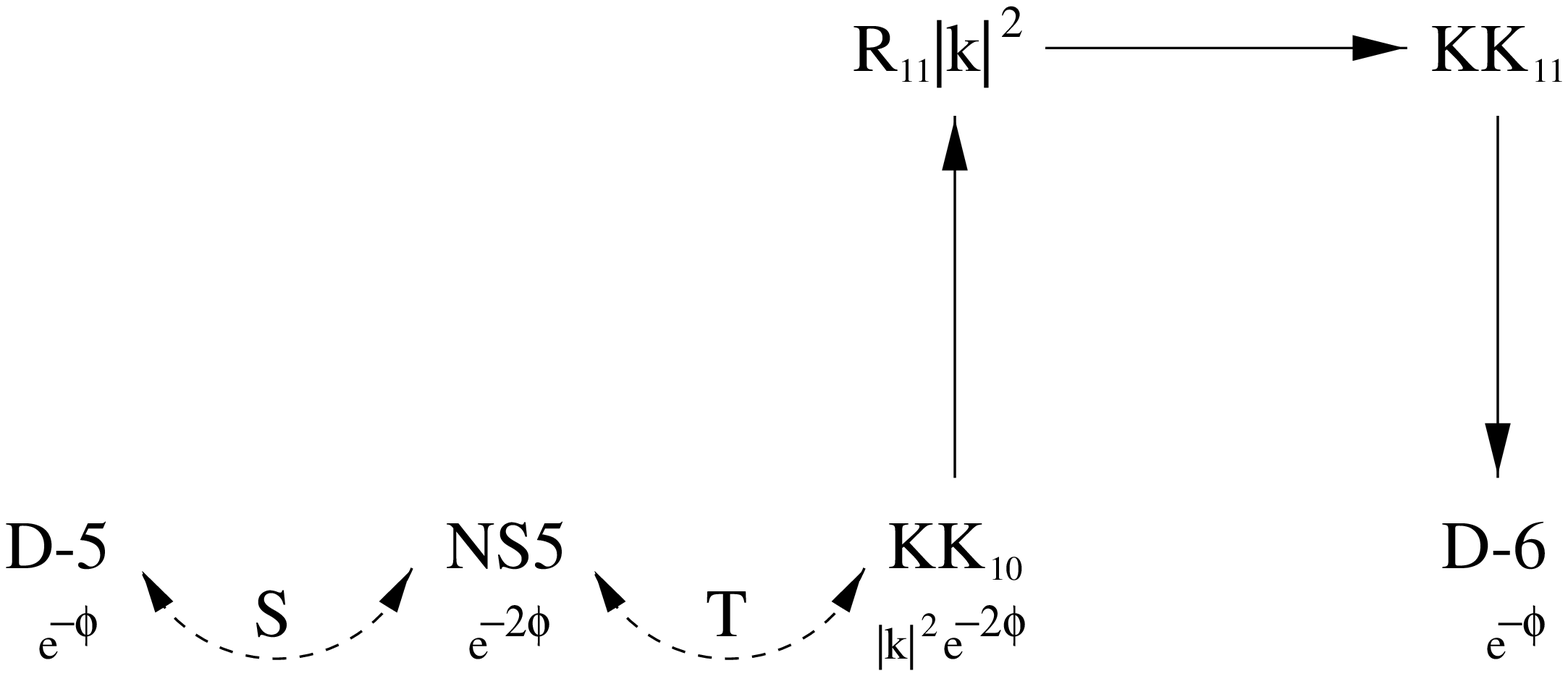}
\end{center}
Figure 3: {\small {\bf The effective Type IIA KK-monopole tension:} 
The arrow
pointing up relates the effective action
 of the IIA KK-monopole and the gauged 
$\sigma$-model action. The arrow to the right relates the gauged 
$\sigma$-model action to the d=11 KK-monopole action and the arrow pointing 
down represents direct dimensional reduction to the D-6-brane action. The 
curved arrows indicate (T and S) duality transformations. Below each brane
we have indicated how the effective tension scales.} 
\end{figure} 

The discussion above is summarized in Figure 3. In this figure we have
also indicated how the effective tension of the d=10 Type IIA KK monopole 
transforms into the effective tension of the Type IIB NS5-brane via
T-duality in the Killing isometry direction. Here we have used the
standard T-duality rules: 

\begin{equation}
\label{Tduality}
|k| \rightarrow {1\over |k|}\, ,\hskip 1.5truecm 
e^\phi \rightarrow {1\over |k|} e^\phi\, .
\end{equation}
Furthermore we have indicated how the effective tension of the Type IIB 
NS5-brane transforms into the effective tension of the Type IIB D-5-brane
under a S-duality transformation. In general the effective tension
of a D-p-brane (p odd) transforms under S-duality as

\begin{equation}
e^{-\phi}\sqrt {|g|} \rightarrow e^{\delta\phi}\sqrt {|g|}
\end{equation}
with

\begin{equation} 
\delta = -{\textstyle{{1\over 2}}} (p-1)\, .
\label{Sdual}
\end{equation}
The case in Figure 3 corresponds to $p=5$. The exact transformations between
the complete worldvolume 5-brane actions is discussed in \cite{preparation}.

We now turn our attention to the M-9-brane. 
We have seen that the oxidation of the D-8-brane leads to a gauged
sigma model that scales as $(R_{11})^3$ (see Figure 1). 
Therefore the M-9-brane cannot
move in eleven decompactified dimensions. At this stage we mimic
the situation for the KK-monopole and propose the following (Nambu-Goto part
of the) worldvolume M-9-brane action:

\begin{equation}
S_{M-9} =-T \int d^{9}\hat{x} \hspace{4pt} |\hat{k}|^{3} 
\sqrt{\mid \partial_i \hat{X}^{\hat{\mu}} \partial_j \hat{X}^{\hat{\nu}}
\hat{\Pi}_{\hat{\mu}\hat{\nu}} \mid}\, .
\label{M9action}
\end{equation}
Note that we do not integrate over the special (compact) isometry direction
defined by the Killing vector. Therefore the action (\ref{M9action})
describes a {\it wrapped} 9-brane instead of a 9-brane with 
9 non-compact (spacelike) worldvolume directions.

By construction, the reduction of the M-9-brane action (\ref{M9action}) 
over the special isometry direction yields
the worldvolume action of the D-8-brane.
Similarly, the direct dimensional reduction 
(i.e.~the reduction  over the single transversal direction) 
is expected to yield the worldvolume action of the conjectured (wrapped)
IIA-9-brane \cite{hull}\footnote{
Note that at the level of solutions the Type IIA 9-brane
is represented by d=10 Minkowski spacetime which has unbroken supersymmetry.
The presence of an orientifold (see below)
breaks half of the supersymmetry.}.
Applying equation (\ref{eqab2}), for 
$(p,\beta,\gamma) = (8,0,3)$\footnote{Note that $\beta$ gives the
scaling with respect to the eleventh direction which in this case is a 
direction transverse to the brane. Therefore we have $\beta=0$.} 
we find $\alpha=-4$. This agrees with the dilaton coupling
predicted in \cite{hull} by other considerations. We conclude that 
the M-9-brane action (\ref{M9action}) 
reproduces the correct dilaton couplings for both the Type IIA D-8-brane as 
well as the Type 
IIA 9-brane upon double and direct dimensional reduction, respectively.

In order to perform the direct dimensional reduction of the M-9-brane
we must compactify the single transverse direction. Charge conservation
requires that after compactification we include another source of charge
such as an orientifold. It would be interesting to see whether, after
including this orientifold, the Type IIA 9-brane leads to a description 
of the $E_8\times E_8$ heterotic superstring
theory. This would provide a T-dual version of the description
of the $SO(32)$ heterotic superstring via Type IIB 9-branes as
discussed in \cite{hull}. In this way every ten-dimensional 9-brane 
corresponds to a $N=1$ superstring theory:

\begin{eqnarray}
{\rm IIA-9} &\leftrightarrow& {\rm heterotic}\ E_8\times E_8\, ,\nonumber\cr
{\rm IIB-9} &\leftrightarrow& {\rm heterotic}\ SO(32)\, ,\cr
{\rm D-9}   &\leftrightarrow& {\rm Type\ I}\ SO(32)\, .\nonumber
\end{eqnarray}

The discussion above is summarized in Figure 4. In this figure we have
also indicated how the effective tension of the wrapped Type IIA 9-brane
transforms into the effective tension of the unwrapped Type IIB 9-brane,
or (1,0)-brane,
under a T-duality transformation. Performing a T-duality in the isometry
direction of the Type IIA 9-brane and applying (\ref{Tduality}) we first
find $|k|^3e^{-4\phi} \rightarrow |k|^{1/2}e^{-4\phi}$ which is the
effective tension of a wrapped Type IIB 9-brane. The factor $|k|^{1/2}$
can now be absorbed into the Nambu-Goto term and we obtain an unwrapped
Type IIB 9-brane. 
We have also indicated how a S-duality transformation relates the
Type IIB 9-brane to the D-9-brane, or (0,1)-brane\footnote{The (1,0)-brane and 
(0,1)-brane are two special cases of a whole family of $(p,q)$ 9-branes
\cite{hull}.}. 
In the latter S-duality transformation
we have used equation (\ref{Sdual}) for $p=9$. 
Note that the T-duality between the wrapped Type IIA 9-brane and the unwrapped
Type IIB 9-brane is similar to the T-duality between the (wrapped)
Type IIA KK monopole and the unwrapped Type IIB NS5-brane. In both cases
one brane has a special Killing vector (Type IIA KK monopole, Type IIA 
9-brane) whereas the T-dual brane has no such Killing
vector (Type IIB NS5-brane, Type IIB 9-brane). Figure 4 shows that S- and
T-duality and the existence of the D-9-brane provide another argument in 
favour of the Type IIA 9-brane and its M-theory origin, the M-9-brane.

\begin{figure} \label{action9}
\begin{center}
\includegraphics[angle=0, width=150mm]{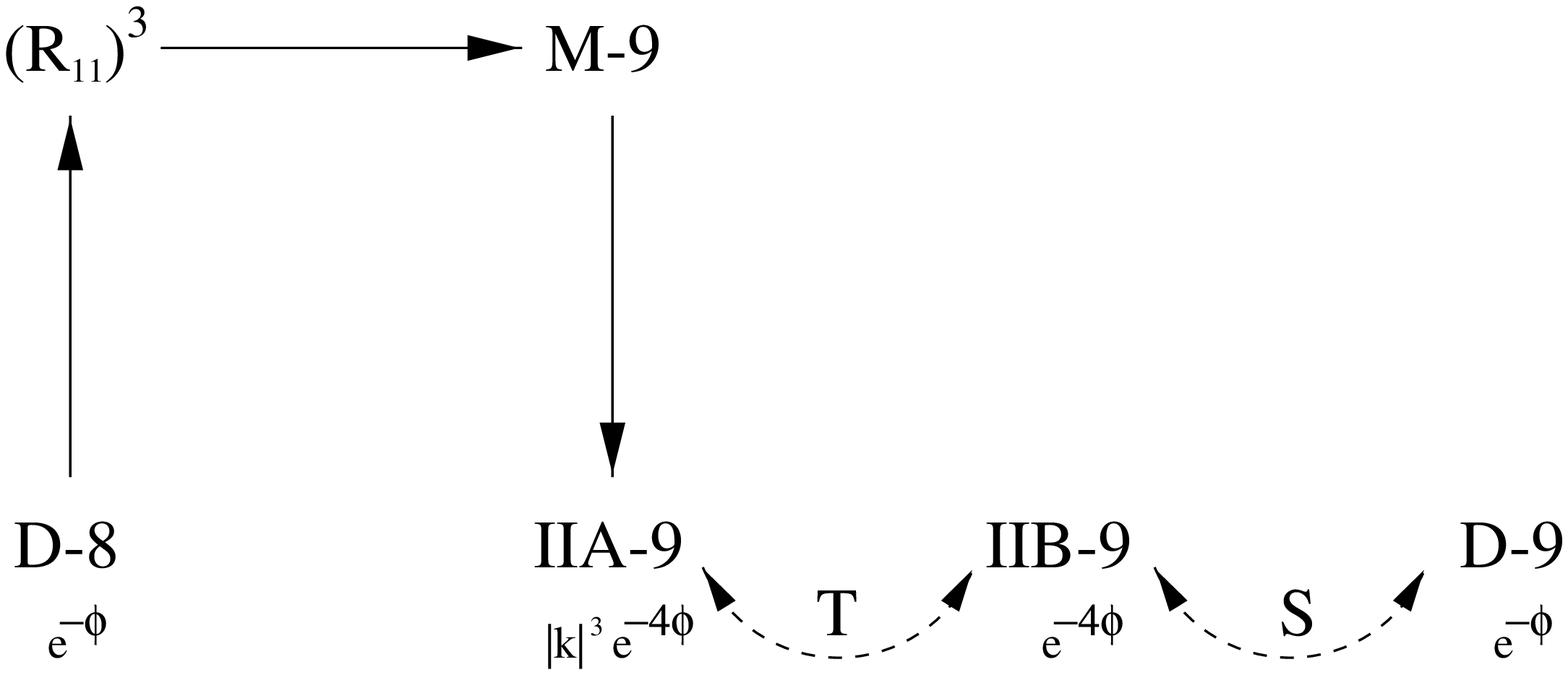}
\end{center}
Figure 4: {\small {\bf The M-9-brane effective tension:} 
The arrow
pointing up relates the effective action
 of the D-8-brane and the gauged 
$\sigma$-model action. The arrow to the right relates the gauged 
$\sigma$-model action to the M-9-brane action and the arrow pointing 
down represents direct dimensional reduction to the (wrapped)
Type IIA 9-brane action. 
The curved arrows indicate (T and S) duality transformations. Below each brane
we have indicated how the effective tension scales.} 
\end{figure}

So far, we have only discussed the leading Nambu-Goto term of the M-9-brane
worldvolume action. In principle one should be able to reconstruct the 
complete action by using the relation, via reduction and duality,
with the worldvolume actions of the other branes given in Figure 4. 
Since we are
dealing with a wrapped M-9-brane we have a 9-dimensional worldvolume
and therefore we expect the worldvolume fields to form a d=9 vector
multiplet. The single scalar of this vector multiplet describes
the dynamics in the single transverse direction in which the M-9-brane
can move. It would be interesting to explicitly construct the complete
M-9-brane worldvolume action.

\vspace{0.5cm}
\noindent{\bf 3.\ The D-8-brane target space solution}
\vspace{0.5cm}

Before discussing the M-9-brane target space solution in the next section
we will first 
review in this section some properties of the D-8-brane target space solution.
The D-8-brane target space solution arises as a domain wall 
solution\footnote{For earlier work on domain wall solutions in d=4
supergravity theories, see \cite{cvetic}.}
in massive IIA supergravity \cite{romans}.
When the cosmological constant term is dualized
to a 9-form potential one can write down a domain wall solution 
with different values
of the cosmological constant at different sides of the domain wall 
\cite{berogrpato1}. This solution is coordinate equivalent
to the conformally flat solution given in \cite{PW}.

We consider the following Ansatz for the extreme D-8-brane solution
(in string frame metric):

\begin{eqnarray}
\label{ansatz}
ds^2_{10} &=& H^{\alpha}(dt^2-dx_{(8)}^2)-H^{\beta} dy^2 , \nonumber \\
e^{2\phi} &=& H^{\gamma}\, ,\hskip 1truecm 
C^{(9)}_{012345678} =  H^{\epsilon}\, ,
\end{eqnarray} 
where $H = H(y)$ is a harmonic function over the single transverse direction
$y$ whose form, in a local neighbourhood, is given by

\begin{equation}
H(y) = c + Q|y|
\end{equation}
in terms of two constants $c$ and $Q$. 
In order to avoid a singularity
at $H=0$ and to obtain a real dilaton, we use the absolute value of $y$ in
the harmonic function and take $c > 0$ and $Q >0$.
The Ansatz (\ref{ansatz}) describes a domain wall
positioned at $y=0$. 

It turns out that the parameter $\epsilon$ can not be
determined by the equations of motion obtained by minimizing the 
action (\ref{C9}). This is in contradistinction to
the D-p-branes with $p<8$ where one finds, for all $p<8$, that $\epsilon = -1$.
Solving the equations of motion obtained from the action (\ref{C9})
we find the following expressions for $\alpha, \beta$, $\gamma$:

\begin{equation}
\label{sol}
\alpha = \tfrac{1}{2} \epsilon\, , \hskip 1.5truecm
\beta = - \tfrac{5}{2} \epsilon - 2\, , \hskip 1.5truecm
\gamma = \tfrac{5}{2} \epsilon\, .
\end{equation}
Furthermore, substituting the solution (\ref{sol}) into
(\ref{pdual}), we find the following relation between $m$ and $Q$:

\begin{equation}
m=\pm \epsilon Q\, .
\end{equation}
Notice that for any fixed $m$ and $Q$ we find two
solutions corresponding to $\epsilon$ and $-\epsilon$.  
For $m=0$ there is a single solution with  $\epsilon = 0$ or, equivalently,
$Q=0$. In this case the metric reduces to that of a flat Minkowski spacetime.

The value of the cosmological constant differs at the
two sides of the domain wall whenever we make different
choices for the  constant $Q$ at the left and right of the domain wall:

\begin{eqnarray}
\label{pc}
H (y) &=& c + Q_L |y|\, ,\hskip 1truecm y < 0 , \nonumber \\
H (y) &=& c + Q_R |y|\, ,\hskip 1truecm y > 0\, .
\end{eqnarray}

The free parameter $\epsilon$ labelling the above D-8-brane
solutions is related to the freedom to perform a coordinate transformation
in $y$ keeping
the solution within the ansatz (\ref{ansatz}).
To show this we first note that at one side of the domain wall
one can always use coordinates such that $y \geq 0$. 
By performing a suitable shift
transformation,  $y'=y+c/Q$, the harmonic function can always be 
written as $H(y')=Qy'$, where $y' \in (c/Q,\infty)$ and $Q>0$. We 
only consider coordinate transformations that keep the transversal 
coordinate $y$ within 
this (positive and infinite) range.  Consider a D-8-brane
for a given negative value of $\epsilon$, say $\epsilon=-1$ or $m=\pm Q$.
We perform
the following coordinate transformation labelled by $\epsilon$ 
(we omit the prime on the shifted coordinate):

\begin{equation}
y \rightarrow y' = f(\epsilon) y^{-\tfrac{1}{\epsilon}}\, , 
\label{transform}
\end{equation}
with the function $f(\epsilon)$ given by

\begin{equation}
f(\epsilon)=-\epsilon Q^{-\tfrac{1+\epsilon}{\epsilon}}.
\end{equation}
We restrict ourselves to negative $\epsilon$ (for positive
$\epsilon$ the range  of $y^\prime$ would become
negative and finite).  
Under the above coordinate transformation
the harmonic function $H$ transforms as
\begin{equation}
H_{(\epsilon=-1)}(y) = -my = \left ( H_{(\epsilon)}(y^\prime)\right )^
{-\epsilon}
= \left (Q^\prime y^\prime\right )^{-\epsilon}\, ,
\end{equation}
with $Q^\prime=- Q/\epsilon$. Therefore
all solutions with negative $\epsilon$, defined at positive $y$, are 
related to each other by the coordinate transformations (\ref{transform}).
The same holds for all positive $\epsilon$
solutions. Starting with a solution for a given positive value of
$\epsilon$, say $\epsilon=1$, we obtain all
other positive $\epsilon$ solutions by performing the transformation
(\ref{transform}) with $\epsilon$ replaced by $-\epsilon$.

Finally, to relate solutions with positive and
negative values of $\epsilon$ one must perform
a coordinate transformation of the form

\begin{equation}
\label{inversion}
y\rightarrow 1/y\, .
\end{equation}
This transformation is included in 
(\ref{transform}) if we also allow positive values of $\epsilon$.
However, under the transformation (\ref{inversion})
the infinite positive domain of $y$ gets mapped to a finite negative
domain in $y^\prime$ and $Q$ becomes negative to keep $H(y)$ positive.
This means that in the new coordinate system
$y=0$ acts as infinity. In any case, we see that for all values of
$\epsilon$, positive and negative, the domain wall solutions are
coordinate equivalent to each other.
In the rest of this work we will often concentrate on the
$\epsilon <0$ solutions because in that case the position of the
domain wall occurs for a finite value of $y$ and the asymptotic region
far away from the domain wall is reached by taking $y \rightarrow \infty$.

We next consider the asymptotic values of the D-8-brane configuration
keeping
in mind that solutions labelled by different values of $\epsilon$ are 
coordinate equivalent. For reasons explained in the previous paragraph
we restrict ourselves to $\epsilon <0$.
We first consider the dilaton.
For $\epsilon<0$, the string coupling $e^{\phi}$ approaches
zero in the limit $y \rightarrow \infty$\footnote{This is different
from the other D-p-branes ($0\le p \le 6$) where the string coupling
approaches a constant at transverse infinity.} and $c^{5\epsilon/4}$ in the
limit $y \rightarrow 0$ (the position of the domain wall). 
Because of the freedom in choosing
$c$ the string coupling can take any value 
at the position of the D-8-brane, when $y\rightarrow 0$. In particular,
it becomes infinite if $c \rightarrow 0$.

To determine the asymptotic structure of the metric we
calculate the Riemann tensor squared. For general (positive and negative)
$\epsilon$ one finds \cite{lupoto}:
\begin{equation}
R_{\mu\nu\delta\rho} R^{\mu\nu\delta\rho} \propto H^{5\epsilon}
(\partial_y H)^4\, .
\end{equation}
This shows that, assuming that $\epsilon<0$, in the limit
$y \rightarrow \infty$ the curvature approaches zero, while the curvature
near the D-8-brane position approaches a constant as long as we keep 
$c>0$. If we take the limit $c \rightarrow 0$ the curvature blows up and 
we are dealing with a true singularity.

It is instructive to consider timelike geodesics in the 
background of a D-8-brane. 
The equation of such a geodesic is given by (in Einstein frame):

\begin{equation}
\ddot{y}= -{1\over 16} \epsilon Q H(y)^{3\epsilon+1} \, .
\label{geodesic}
\end{equation}
We deduce that for $\epsilon<0$ 
test-objects will be repelled by the D-8-brane. Only when $\epsilon<-1/3$
the acceleration will approach zero when $y \rightarrow \infty$.
We can solve (\ref{geodesic}) for general $\epsilon$
and find that, when a 
test-particle is inserted in a D-8-brane geometry with zero velocity,
the geodesic trajectory is described by
\begin{equation}
y(t)=(at^2+b)^{-1/3\epsilon} -c \, ,
\end{equation}
where $a>0$ and $b$ are constants determined by $y(0), Q$ and $m^2$ and $c$ 
is the constant in the harmonic function $H(y)=c + Q|y|$.
This formula shows the difference between the $\epsilon>0$ and 
$\epsilon<0$ solutions.
For positive $\epsilon$ the geodesics approach a finite value of
$y$ if we take $t$ to infinity.
For negative $\epsilon$  the test-particles move towards infinity
if we take $t$ to infinity.  

Finally, for special values of $\epsilon<0$ the D-8-brane solution reduces to
a more familiar form. We mention the following three choices:

\begin{itemize}
\item
$\epsilon=-1$ gives 
\begin{eqnarray}
ds^2_{10} &=& H^{-1/2}(dt^2-dx_{(8)}^2)-H^{1/2} dy^2\, , \nonumber \\
e^{2\phi} &=& H^{-5/2}\, ,\hskip 1truecm 
C^{(9)}_{012345678} =  H^{-1}\, .
\end{eqnarray} 
This is the standard
form of the D-8-brane solution as it is obtained via T-duality from the
other D-p-brane solutions.

\item
$\epsilon=- \tfrac{2}{3}$ gives 
\begin{eqnarray}
ds^2_{10} &=& H^{-1/3}(dt^2-dx_{(8)}^2 - dy^2)\, , \nonumber \\
e^{2\phi} &=& H^{-5/3}\, ,\hskip 1truecm 
C^{(9)}_{012345678} =  H^{-2/3}\, .
\end{eqnarray} 
This is the conformal flat metric solution given in \cite{PW}.

\item
$\epsilon=- \tfrac{4}{5}$ gives
\begin{eqnarray}
ds^2_{10} &=& H^{-2/5}(dt^2-dx_{(8)}^2) - dy^2\, , \nonumber \\
e^{2\phi} &=& H^{-2}\, ,\hskip 1truecm 
C^{(9)}_{012345678} =  H^{-4/5}\, .
\end{eqnarray} 
The special feature of this choice of $\epsilon$ is that
$\beta = 0$.
\end{itemize}

\vspace{0.5cm}
\noindent{\bf 4.\ The M-9-brane target space solution}
\vspace{0.5cm}

In order to oxidize the D-8-brane to an M-theory solution we must add
a cosmological constant, proportional to $m^2$, to the d=11 supergravity
theory. As discussed in the introduction, to do this we must assume
that the d=11 background fields have an isometry generated by a Killing
vector $k^\mu$ (in this section we omit hats on the fields).
The extra eleventh direction corresponds to this isometry direction.
Notice that the present situation differs from the oxidation of the
D-6-brane to an M-theory Kaluza-Klein monopole. In the case of the
monopole the isometry direction in the Taub-NUT space is required in
order to solve the equations of motion of massless d=11 supergravity.
Here, the extra isometry direction is already needed, before solving
the equations of motion, in order to write down the massive d=11 
supergravity Lagrangian.

Although a d=11 manifold with an isometry is basically a d=10
manifold, the d=10 interpretation of a solution may differ from the d=11
interpretation of the same solution. An example of this is the d=11 KK
monopole. Its d=10 interpretation is the D-6-brane which is singular.
The d=11 interpretation of the same solution leads to a non-singular solution.
In order to show the regularity of the solution one crucially needs the
compact isometry direction.
The singularity of the ten-dimensional D-6-brane can be viewed as an 
illegitimate 
neglect of KK modes wich become massless at the D-6-brane core
(see e.g.~\cite{townsend}). 
In the case of
the M-9-brane the situation is different since, in order to write down
an eleven-dimensional action, we have already restricted ourselves to
background fields with an isometry.

Assuming the extra isometry direction it is straightforward 
to rewrite the D-8-brane metric and dilaton 
in terms of the M-theory metric. We use
the following reduction ansatz for the d=11 metric:

\begin{equation}
ds_{11}^2=e^{-\tfrac{2}{3}\phi} ds_{10}^2  + e^{\tfrac{4}{3}\phi}dz^2\, ,
\end{equation}
where $z$ is the isometry direction and $ds_{10}^2$ is the
ten-dimensional string frame metric. Applying this formula we obtain
the following oxidized D-8-brane, or (wrapped) M-9-brane, solution:

\begin{eqnarray}
\label{sol11}
ds_{11}^2 &=& H^{-\tfrac{1}{3}\epsilon} \left( dt^2 - dx_{(8)}^2 \right)
- H^{-\tfrac{10}{3}\epsilon -2} dy^2 
-H^{\tfrac{5}{3}\epsilon} dz^2\, , \nonumber \\
\\
H(y) &=& c + Q|y| \hskip .5truecm (c,Q > 0)\, ,\hskip1.5truecm 
m = \pm \epsilon Q\, . \nonumber
\end{eqnarray}
This M-9-brane solution, when reduced over the $z$-direction, 
gives the D-8-brane solution of the previous section. The arbitrary
parameter $\epsilon$ is related to a coordinate transformation
in $y$, as explained in the previous section. 

For generic values of $\epsilon$ the M-9-brane metric given in 
(\ref{sol11}) represents a 3-block solution, i.e.~it exhibits
three inequivalent directions. This is different from the M-2-brane
and M-5-brane which are represented by 2-block solutions with the
two inequivalent directions corresponding to the worldvolume and transverse
directions. Only for special values of $\epsilon$ the M-9-brane metric 
also reduces to a 2-block solution:

\begin{itemize}
\item $\epsilon = - \tfrac{2}{5}$ gives

\begin{equation}
ds^2_{11} = H^{2/15} (dt^2 - dx^2_{(8)}) - H^{-2/3} (dy^2 + dz^2)\, .
\end{equation}

\item $\epsilon = - \tfrac{2}{3}$ gives

\begin{equation}
ds^2_{11} = H^{2/9} (dt^2 - dx^2_{(8)} - dy^2) - H^{-10/9}dz^2\, .
\end{equation}
\end{itemize}

\noindent The first case gives a metric suggesting 
$z$ represents a special isometry direction transverse to the brane.
This is what happens for the M-theory Kaluza-Klein monopole.

We first consider in more detail the properties of a single M-9-brane. Next, we
will study a system of two M-9-branes.
\bigskip

\noindent {\it The Single M-9-brane}
\bigskip

We consider a single domain wall solution, positioned at $y=0$,
with harmonic function given by (\ref{pc}).
Since $g_{zz}= (R_{11})^2$ it follows from the M-9-brane metric that
$R_{11}=H(y)^{\tfrac{5}{6}\epsilon}$. We take $\epsilon <0$ for 
reasons explained in the previous section.
We conclude that for an observer far away from the M-9-brane, spacetime 
is effectively ten-dimensional: $R_{11}\rightarrow 0$ if $y \rightarrow 
\infty$. On the other hand, close to the M-9-brane
position $R_{11}$ approaches a constant value: $R_{11}\rightarrow 
c^{\tfrac{5}{6}\epsilon}$ if $y \rightarrow 0$. For a fixed (negative)
value of $\epsilon$ the radius at the position of the domain wall 
becomes very large if the value of $c$ is chosen to be small: 
$R_{11}\rightarrow \infty$
if $c\rightarrow 0$\footnote{Note that the curvature at $y=0$ becomes
infinite in the limit $c\rightarrow 0$.}. Thus we see that 
for a general M-9-brane
the asymptotic geometry ($|y| \rightarrow \infty$) looks ten-dimensional 
and only when we 
approach the domain wall the eleventh dimension opens up. 
 
\begin{figure} \label{fig3}
\begin{center}
\includegraphics[angle=0, width=150mm]{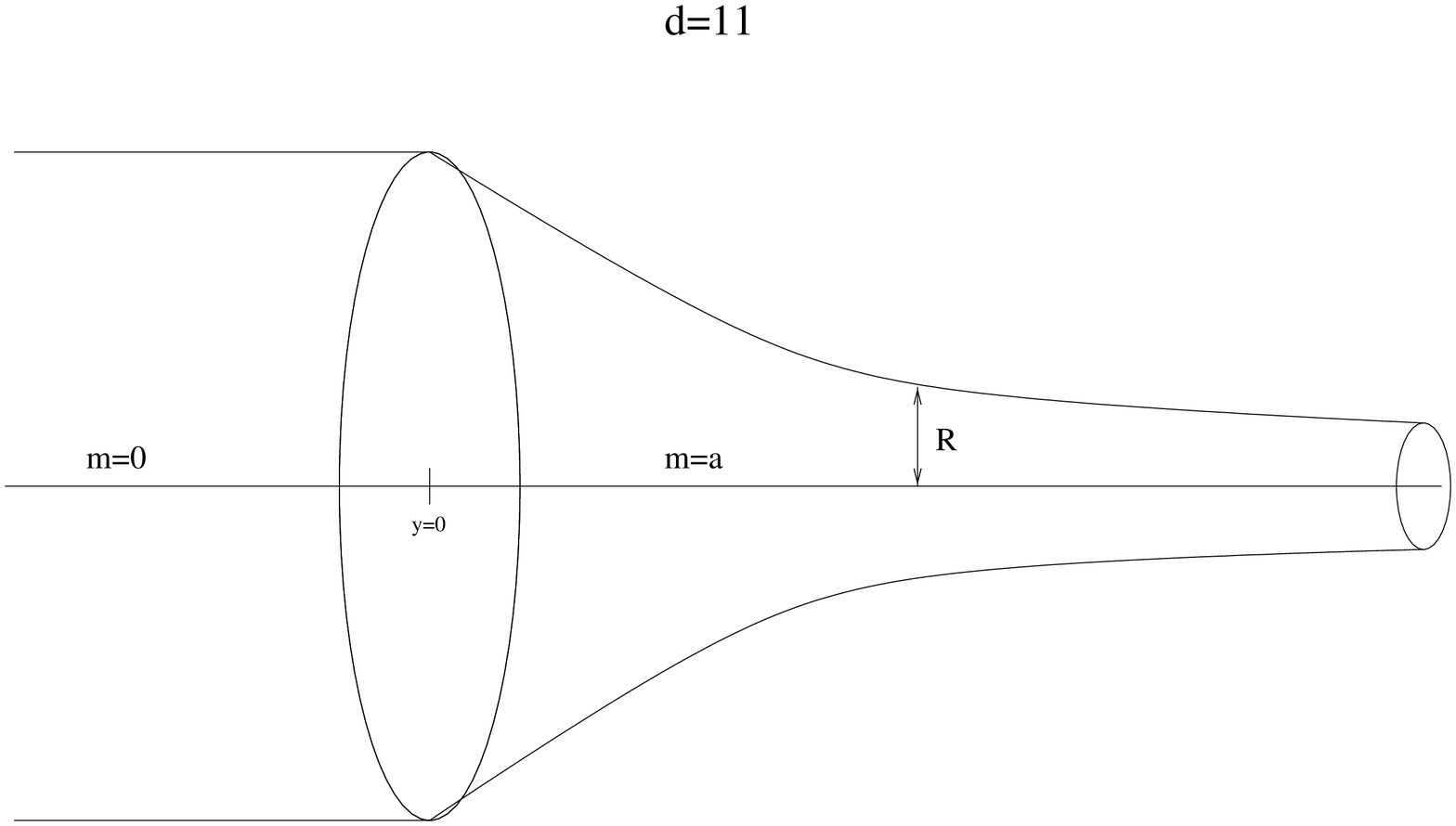}
\end{center}
Figure 5: {\small {\bf The single M-9-brane:} The cosmological constant 
has been taken equal to zero at the left of the domain wall and 
non-zero at the right. 
For $c\ne 0\ (c\rightarrow 0)$ the radius of the eleventh
dimension $R_{11}$ is finite (infinite) at $y=0$. In the limit
$c\rightarrow 0$ the asymptotic 
spacetime is given by
ten-dimensional (eleven-dimensional) Minkowski space for $y\rightarrow 
\infty\ ( y \rightarrow -\infty)$.}
\end{figure}

We now consider
a special case in which the cosmological constant is zero at one
side of the M-9-brane (say for $y<0$), i.e.

\begin{eqnarray}
 H(y) &=& c \hskip 2.5truecm  y<0\, ,\nonumber\\
 H(y) &=& c + Q_R |y|\hskip 1truecm  y>0\, .
\end{eqnarray}
In that case the d=11 M-theory metric is given by 

\begin{eqnarray}
ds_{11}^2 &=& c^{-\tfrac{1}{3}\epsilon} \left( dt^2 - dx_{(8)}^2 \right)
- c^{-\tfrac{10}{3}\epsilon -2} dy^2 
-c^{\tfrac{5}{3}\epsilon} dz^2\, , \hskip 1truecm y < 0\, ,\nonumber\\
ds_{11}^2 &=& H^{-\tfrac{1}{3}\epsilon} \left( dt^2 - dx_{(8)}^2 \right)
- H^{-\tfrac{10}{3}\epsilon -2} dy^2 
-H^{\tfrac{5}{3}\epsilon} dz^2\, , \hskip .5truecm y >0\, .
\end{eqnarray}
Thus, we see that at the $y<0$ side of the domain wall, the d=11 spacetime
is given by the direct product of a d=10 Minkowski spacetime and a circle
with radius $R_{11} = c^{5\epsilon/6}$. Since $\epsilon <0$
we obtain in the limit $c\rightarrow 0$ an {\it unwrapped}
 domain wall whose geometry at
one side is given by a decompactified d=11 Minkowski spacetime and
at the other side by a d=11 geometry with an isometry which, far away
from the domain wall, looks like a ten-dimensional spacetime (see Figure 5).
The domain wall interpolates between two spacetimes which are related
to each other by a decompactification of the special $z$ direction.

\bigskip
\noindent {\it Two M-9-branes}
\bigskip

We next consider two parallel
M-9-branes where each of the two domain walls have
the product space $M^{10} \times S^1$ at one side, as described above. 
To obtain  a static solution the two M-9-branes must have the same ``charge''.
In order for this to be the case the
$M^{10} \times S^1$ spacetime must lie in between the two 
M-9-branes and the cosmological constants at the other side of the
two M-9-branes must be equal to each other (see Figure 6).  To be precise
\begin{eqnarray}
H_1 (y) &=& Q |y-y_1| + c\, ,\hskip 1truecm H_2(y) = c\, , \hskip 1.2truecm
 \nonumber  y < y_1\, ,\nonumber \\
H_1 (y) &=&  H_2(y) = c\, , \hskip 3.9truecm y_1 < y < y_2\, ,  \\
H_1(y) &=&  c\, ,\hskip .5truecm H_2(y) = Q|y - y_2|\, ,\hskip 2.4truecm
y > y_2\, , \nonumber
\end{eqnarray}
where $y= y_1 \ (y_2)$ is the position of the first (second) 
domain wall.

Consider the region in between the two M-9-branes. In this region
we can take the limit $R_{11} \rightarrow \infty$ because 
$m=0$. This allows us to {\it unwrap} the two M-9-branes
and to define a massless bulk supergravity theory 
in between the two M-9-branes. In this way we obtain a system that is
reminiscent to the nine-branes of Ho{\u r}ava and Witten \cite{howi}.

\begin{figure} \label{fig4}
\begin{center}
\includegraphics[angle=0, width=150mm]{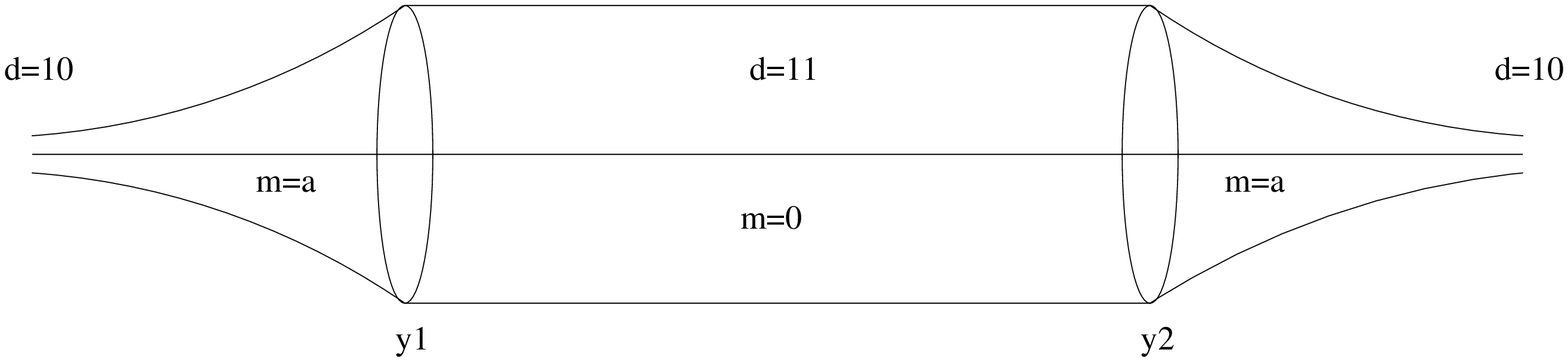}
\end{center}
Figure 6: {\small {\bf Two M-9-branes:} The radius of the eleventh
dimension $R_{11}$ is finite in between the two domain walls. In this region
spacetime
can be decompactified in the $z$ direction
by taking the limit $c\rightarrow 0$ or
$R_{11} \rightarrow \infty$.}
\end{figure}

In \cite{howi} eleven-dimensional
supergravity is considered on the manifold with boundary ${\bf R}^{10} \times
{\bf S}^1/{\bf Z}_2 = {\bf R}^{10} \times I$ with $I$ the unit interval.
The boundary of this manifold consists of two ``end of the world'' nine
branes. Anomaly considerations determine that the worldvolume theory of each
nine brane is given by a N=1 vector multiplet with gauge group $E_8$.
The bulk theory in between the two nine-branes is given by massless
d=11 supergravity. According to \cite{howi} the strong coupling
limit of the $E_8\times E_8$ heterotic string is given by massless
d=11 supergravity
on $R^{10} \times I$ in the same way as the strong coupling
behaviour of the Type IIA superstring is given by massless
d=11 supergravity on $R^{10}
\times S^1$. In this approach the heterotic dilaton is related to the
radius $R_{11}$ in the $y$-direction\footnote{Note that in the double
dimensional reduction leading to the D-8-brane the 
IIA dilaton is related to the radius $R_{11}$ in the $z$-direction.}.
In the limit $R_{11}\rightarrow 0$ the two nine-branes coincide and we
end up with a d=10 spacetime. 

In the Conclusions we will
discuss the similarities and differences between the two M-9-brane system
and the Ho\v rava-Witten system sketched above.

\vspace{0.5cm}
\noindent{\bf 5.\ Supersymmetry}
\vspace{0.5cm}

We first consider the supersymmetry preserved by the D-8-brane.
The relevant part of the supersymmetry rules (with parameter $\eta$)
of massive IIA supergravity is given by (in string frame) \cite{susyrules}

\begin{eqnarray}
\delta \psi_\mu &=& \partial_\mu\eta - \tfrac{1}{4}\omega_\mu{}^{ab}
\Gamma_{ab}\eta + \tfrac{1}{8} m e^\phi \Gamma_{\star} \Gamma_\mu\eta\, ,
\nonumber\\
\delta \lambda &=& \Gamma^\mu (\partial_\mu\phi)\eta - \tfrac{5}{4}
m e^\phi  \Gamma_{\star} \eta\, . 
\label{10susy}
\end{eqnarray}
Substituting the D-8-brane solution (\ref{ansatz}) 
into the supersymmetry preserving 
conditions $\delta \psi_\mu = \delta\lambda = 0$ we find that the
D-8-brane, for each value of $\epsilon$, breaks half of the bulk
supersymmetry, with the Killing spinor given by \cite{PW,berogrpato1}

\begin{equation}
(1 + \Gamma_{012345678})\eta = 0\, ,\hskip 1.5truecm 
\eta = H^{\epsilon/8}\eta_0
\label{susy}
\end{equation}
for constant $\eta_0$.

Substituting the D-8-brane solution into the
supersymmetry transformation rules (\ref{10susy}) we find an overall factor 
$H(y)^{\frac{5\epsilon}{4}}$.
For negative $\epsilon$ we 
see that in the limit $y\rightarrow \infty$ we have unbroken supersymmetry.
We do not find supersymmetry enhancement in the other asymptotic
region, i.e.~at the position of the domain wall. 

We next consider the supersymmetry preserved by the M-9-brane. The
d=10 massive supersymmetry rules given in equation (\ref{10susy}) correspond to
the following supersymmetry rule of the d=11 gravitino (we omit hats):

\begin{equation}
\delta \psi_\mu = \partial_\mu\eta - \tfrac{1}{4}\omega_\mu{}^{ab}
\Gamma_{ab}\eta + \tfrac{i}{24} m |k| (2 k^\nu \Gamma_\nu \Gamma_\mu
-3 k_{\mu}) \eta\, .
\label{11susy}
\end{equation}
We substitute the M-9-brane solution (\ref{sol11})
into the supersymmetry preserving condition $\delta\psi_\mu = 0$
and find the conditions

\begin{equation}
(1 + i\Gamma_z \Gamma_y) \eta = 0\, , \hskip 1.5truecm 
\eta= H^{-\epsilon/12} \eta_0
\end{equation}
for constant $\eta_0$. Assuming that we reduce over the $z$-direction
we have $\Gamma_z = i\Gamma_\star$.
The projection operator $i\Gamma_\star \Gamma_y$ is equivalent 
to the projection operator $\Gamma_{01 \ldots 8}$ which is not yet
of the form (\ref{susycondition}). However, after performing
the redefinition $\Gamma_{\mu} \rightarrow 
\Gamma_{\mu}'=i \Gamma_{\star} \Gamma_{\mu}$\footnote{Using the Killing
vector this redefinition can also be written in an eleven-dimensional
notation.}, we obtain

\begin{equation}
(1 + \Gamma_{01 \ldots 8z}) \eta = 0\, ,
\end{equation}
which is exactly the projection operator of an M-9-brane as suggested by the
M-theory supersymmetry algebra, see equation (\ref{susycondition}).

\vspace{0.5cm}
\noindent{\bf 6.\ Conclusions}
\vspace{0.5cm}

In this work we have investigated some properties of the M-9-brane
whose existence is suggested by the structure of the M-theory superalgebra
\cite{townsend,hull}. We have studied the M-9-brane from the point of
view of both the worldvolume action as well as the target space solution.
We have shown how the action and solution are related, via reduction and
duality, to the actions and solutions of d=10 superstring theory.

We expect that the
complete M-9-brane action can be constructed by making use of the
connection with the known D-9-brane action.
Concerning the target space solutions we note that there are
several similarities but also differences between the two M-9-brane
system considered in this work and the two nine-brane system of 
Ho\v rava-Witten \cite{howi}. 
One similarity is that the d=10 worldvolume 
theory of both the M-9-brane and the nine-brane of \cite{howi} have
N=1 chiral supersymmetry. Another suggestive similarity is that intersections
of an M-9-brane with one of the other branes of M-theory (see Figure 1)
leads to only 1-branes and 5-branes which are exactly the basic objects in 
$E_8\times E_8$ heterotic superstring theory \cite{BT}. The possible
intersections are \cite{mees}:

\begin{equation}
(1|W,M9)\, ,\hskip .5truecm
(1|M2,M9)\, ,\hskip .5truecm
(5|M5,M9)\, ,\hskip .5truecm
(5|KK,M9)\, .
\end{equation}

A difference between our two M-9-brane system and the one of \cite{howi}
is that in our case the two-domain wall system arises as a solution
to the equations of motion of massive d=11 supergravity whereas in
\cite{howi} the same configuration is taken as a fixed background for
the massless supergravity theory. In our case the d=11 spacetime extends
behind  the domain walls (see Figure 6), whereas the 9-branes of \cite{howi}
are ``one-sided''  9-branes positioned at the end of the 
world\footnote{Recently, one-sided
domain wall solutions have been investigated (G.~Gibbons, private 
communication).}.

Another, related, difference is that in our case, as we have shown
in Section 5,
the d=11 bulk supersymmetry is broken by the M-9-brane solution. On the
other hand, in the two nine-brane system of \cite{howi} supersymmetry
is broken by the boundary conditions on the eleven-dimensional fields. 

It remains to be further investigated whether or not the two M-9-brane
system we have considered in this work is related to the two
``end of the world'' nine-branes of Ho{\u r}ava and Witten.
We hope to report on progress in this direction in the near future.

\vspace{0.5cm}
\noindent{\bf Acknowledgements}
\bigskip

We would like to thank M.~Duff, G.~Gibbons,
M.~Green, P.~Howe, C.~Hull, Y.~Lozano, D.~O'Driscoll,
S.-J.~Rey, M.~de~Roo  and especially E.~Eyras for useful discussions.
The work of J.P.~v.d.~S. is part of the research program of the 
``Stichting voor Fundamenteel Onderzoek der Materie'' (FOM).

\end{document}